\documentclass[twocolumn]{aastex63}
\usepackage{amssymb,graphicx}
\usepackage{natbib}
\setcitestyle{notesep={ }} 
\newcommand{\dv}{\ensuremath{\Delta v_{\rm los}}}
\newcommand{\dproj}{\ensuremath{d_{\rm proj}}}
\newcommand{\Chandra}{\textit{Chandra}}
\newcommand{\XMM}{{XMM-\textit{Newton}}}
\newcommand{\RM}{{redMaPPer}}
\newcommand{\arcs}{\ensuremath{^{\prime\prime}}}
\newcommand{\arcm}{\ensuremath{^{\prime}}}
\newcommand{\sDM}{\ensuremath{\frac{\sigma_{\rm DM}}{m}}}
\begin{document}

\title{A New Galaxy Cluster Merger Capable of Probing Dark Matter:
  Abell 56}

\shorttitle{A New Galaxy Cluster Merger: Abell 56}

\author[0000-0002-0813-5888]{David Wittman}
\affiliation{Department of Physics and Astronomy, University of California, Davis, CA 
  95616 USA}
\author[0000-0002-6217-4861]{Rodrigo Stancioli}
\affiliation{Department of Physics and Astronomy, University of California, Davis, CA 
  95616 USA}
\author[0000-0002-4462-0709]{Kyle Finner}
\affiliation{IPAC, California Institute of Technology, 1200 E California Blvd., Pasadena, CA 91125, USA}
\author[0009-0007-5074-5595]{Faik Bouhrik}
\affiliation{California Northstate University, 2910 Prospect Park Dr,
  Rancho Cordova, CA 95670 USA}
\author[0000-0002-0587-1660]{Reinout van Weeren}\affiliation{Leiden Observatory, Leiden University,
Niels Bohrweg 2, 2333 CA, Leiden, the Netherlands}
\author[0000-0002-9325-1567]{Andrea Botteon}\affiliation{INAF - IRA, via P. Gobetti 101, I-40129 Bologna, Italy}


\begin{abstract} 
  We report the discovery of a binary galaxy cluster merger via a
  search of the redMaPPer optical cluster catalog, with a projected
  separation of 535 kpc between the BCGs.  Archival
  XMM-\textit{Newton} spectro-imaging reveals a gas peak between the
  BCGs, suggesting a recent pericenter passage. We conduct a galaxy
  redshift survey to quantify the line-of-sight velocity difference
  ($153\pm281$ km/s) between the two subclusters.  We present weak
  lensing mass maps from archival HST/ACS imaging, revealing masses of
  $M_{200}=4.5\pm0.8\times10^{14}$ and $2.8\pm0.7\times10^{14}$
  M$_\odot$ associated with the southern and northern galaxy
  subclusters respectively. We also present deep GMRT 650 MHz data
  revealing extended emission, 420 kpc long, which may be an AGN tail
  but is potentially also a candidate radio relic.  We draw from
  cosmological n-body simulations to find analog systems, which imply that
  this system is observed fairly soon (60-271 Myr) after pericenter,
  and that the subcluster separation vector is within 22$^\circ$ of
  the plane of the sky, making it suitable for an estimate of the dark
  matter scattering cross section. We find $\sDM=1.1\pm0.6$ cm$^2$/g,
  suggesting that further study of this system could support
  interestingly tight constraints.
\end{abstract}

\section{Introduction}\label{sec-intro}


A collision of two galaxy clusters dramatically reveals the
contrasting behaviors of gas, galaxies, and dark matter (DM). Seminal
papers on the Bullet Cluster provided a ``direct empirical proof of
dark matter'' \citep{Clowe06} as well as limits on the scattering
cross-section of DM particles with each other
\citep{Markevitch04,Randall2008}, aka DM ``self-interaction.''  The
Bullet constraint, $\sDM< 0.7$ cm$^2$/g, is still quite
large in particle physics terms---roughly at the level of
neutron-neutron scattering.  In principle, ensembles of merging
clusters enable tighter constraints \citep{Harvey15,Wittman18SIDM} but
these are complicated by the fact that few systems have well-modeled
dynamics. Specifically, the time since pericenter, pericenter speed,
and viewing angle cannot be extracted from systems that have more than
two merging subclusters. Even with binary mergers, other factors may
hinder the study of dark matter, such as a merger axis closer to the
line of sight rather than the plane of the sky or bright stars that
limit deep optical observations.
Hence there is interest in finding more ``clean'' binary systems with
merger axis close to the plane of the sky.

\begin{figure*}
\centerline{\includegraphics[width=3.4in]{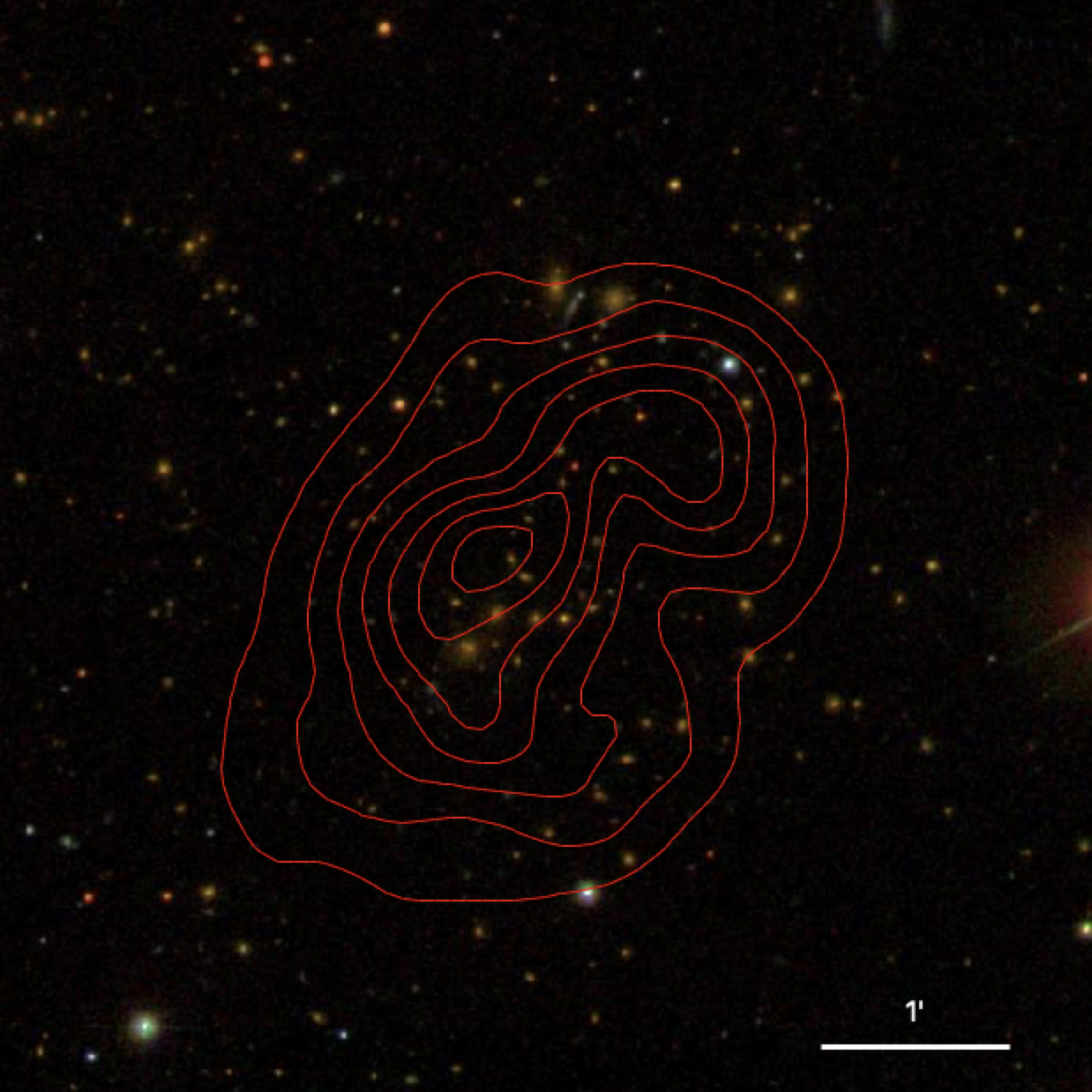}\hskip3mm\includegraphics[width=3.6in]{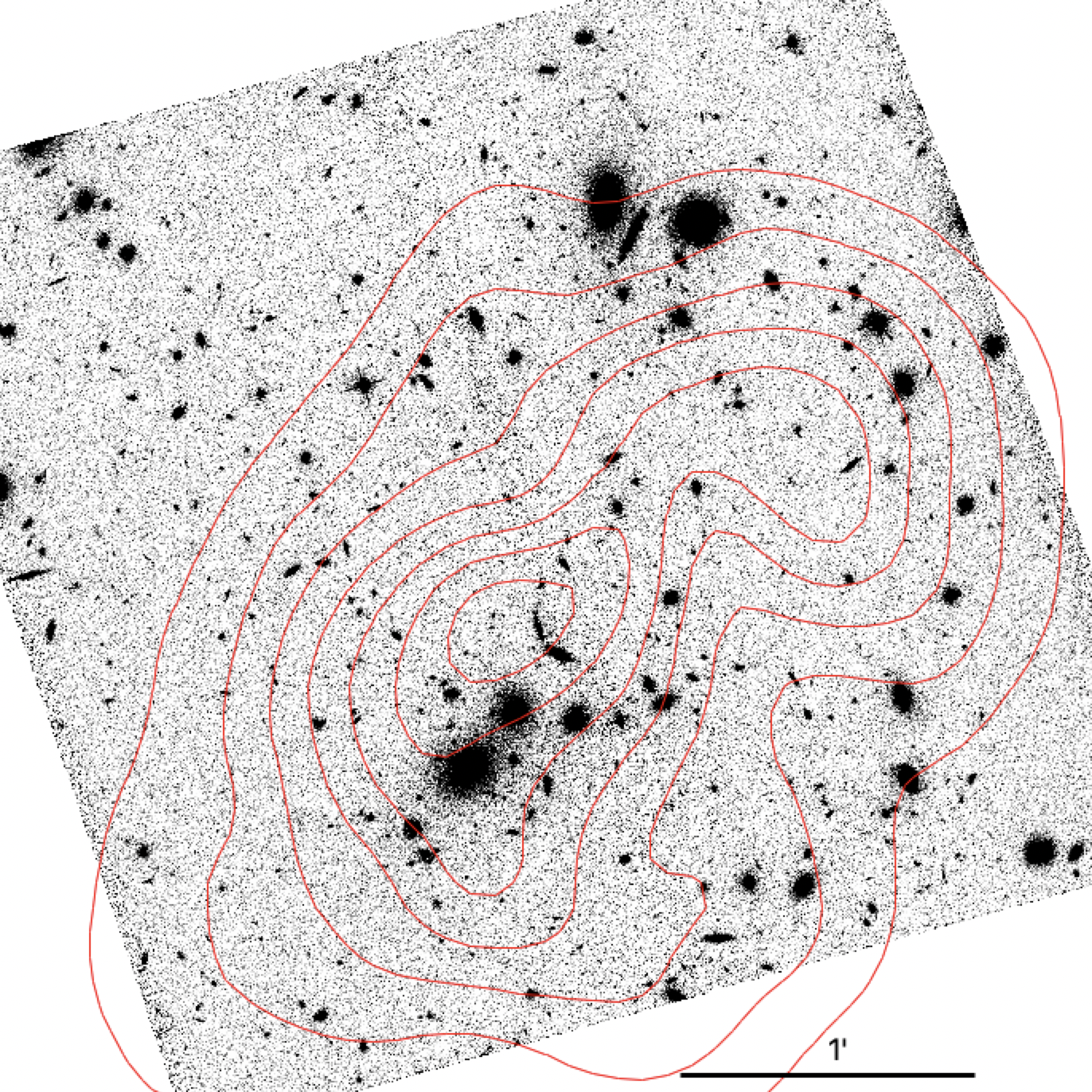}}
\caption{Abell 56: 0.4-1.25 keV \XMM\ contours over SDSS multiband (left) and 
HST/ACS F814W (right) images.}
\label{fig-ovw}
\end{figure*}

Historically, merging systems were discovered upon notice of disturbed
X-ray morphology, which typically happened serendipitously in pointed
observations.  Meanwhile, modern optical sky surveys find tens of
thousands of clusters and promise to find many more as they get wider
and deeper \citep{Euclid2016,LSST2019}.  These surveys potentially
contain new binary mergers, if appropriate cuts can filter out tens of
thousands of more ordinary clusters.  We have developed a new
selection method based on the redMaPPer \citep{Rykoff2014} cluster
catalog, which is in turn based on the ${\sim}$10,000 deg$^2$ Sloan
Digital Sky Survey \citep[SDSS;][]{SDSS2000} imaging. We select
clusters not dominated by a single brightest cluster galaxy (BCG), in
which there is substantial angular separation between the top BCG
candidates. These clusters become candidate mergers, which are then
checked against archival \XMM\ and \Chandra\ data where available; if
the X-ray peak is between the BCGs, the candidate is worthy of
additional followup.  Because these two X-ray archives consist of
pointed observations rather than a uniform survey, our initial
candidates do not form a sample with well-defined selection
criteria. Nevertheless, a few initial candidates are worthy of
immediate study in their own right. In this paper we present the first
such candidate, RM J003353.1-075210.4, which we identify as Abell 56
as explained in \S\ref{sec-overview}. Additional sections in this
paper present a galaxy redshift survey of the system (\S\ref{sec-z});
a weak lensing analysis (\S\ref{sec-wl}); a search for analog systems
in a cosmological simulation (\S\ref{sec-analogs}); radio observations
in search of a radio relic that could outline a merger shock
(\S\ref{sec-GMRT}); and a constraint on the dark matter scattering
cross section \sDM (\S\ref{sec-DM}). We assume a flat $\Lambda$CDM cosmology with
$H_0=69.6$ km/s and $\Omega_m=0.286$.


\section{Abell 56: initial overview}\label{sec-overview}

\textit{Nomenclature.} The redMaPPer designation for this cluster is
RM J003353.1-075210.4. The original coordinates for Abell 56
\citep[][; hereafter ACO]{ACO1989} are nearly $5^\prime$ north of the
redMaPPer position.  Although ACO cite the positional uncertainty as
$2.5^\prime$, inspection of the SDSS
imaging\footnote{\url{https://skyserver.sdss.org/dr16/en/tools/chart/navi.aspx}}
reveals no other clusters in the area, suggesting that the ACO
coordinates are off by more than their nominal uncertainty.  (The
limited depth of the ACO catalog is such that any real ACO cluster
must be in the \RM\ catalog.)  Indeed, the widely-used SIMBAD
database\footnote{\url{http://simbad.u-strasbg.fr/simbad/}} resolves the
name ``Abell 56'' to the redMaPPer position, as does the SDSS
Navigator noted above.  Hence we identify this cluster as Abell 56.
Note, however, that the NASA/IPAC Extragalactic
Database \citep{NEDDOI} resolves this
name to the original ACO coordinates.

This cluster has also been detected by the {\it Planck}
Sunyaev-Zel'dovich survey \citep{PSZ2}, with the designation PSZ2
G109.99-70.28.  This gas peak position is $1.3\arcm\pm2.4\arcm$ from
the redMaPPer position and $4.0\arcm\pm2.4\arcm$ from the ACO
position.  Hence \citet{PSZ2} adopted the redMaPPer position while
adopting ``ACO 56'' as the identifier in their union catalog.  As a
result, a search for G109.99-70.28 on SIMBAD yields the redMaPPer
position; however NED yields the much more uncertain gas peak
position.

\textit{BCGs and redshifts.}  Figure~\ref{fig-ovw} presents two views
of Abell 56: with \XMM\ contours (see below) over SDSS multiband
imaging and over a single-band (F814W) archival image from the Hubble
Space Telescope Advanced Camera for Surveys (HST/ACS) (see
\S\ref{sec-wl}).  At the \RM\ photometric redshift of 0.30, the
physical scale is 4.5 kpc/arcsec. There are two galaxy subclusters
separated by close to $2^\prime$ (530 kpc), with the X-ray peak
located along the subcluster separation vector, about
$32^{\prime\prime}$ (140 kpc) from the southern subcluster. These
numbers will be refined with further data in later sections of this
work.

The southern subcluster is dominated by a galaxy observed by the
Baryon Oscillation Spectroscopic Survey
\citep[BOSS;][]{BOSSoverview2013} to be at $z=0.30231$; redMaPPer
assigns this galaxy 84\% probability of being the overall BCG.  The
northern subcluster has two galaxies that appear nearly equally
salient in Figure~\ref{fig-ovw}; the eastern one (about 0.2 mag
brighter) is assigned 16\% probability of being the overall BCG and
BOSS places it at $z=0.30475$. \RM\ technically assigns some nonzero
BCG probability to the second-brightest galaxy in each subcluster, but
in each case it is only $0.017\%$, which we consider negligible.

\citet{MACS-HST2018} briefly considered this cluster as part of an
86-cluster sample. It is classified in their Table~6 as being in the
most disturbed of their four optical morphology classes.
  
\textit{Merger basics.} Taking the BCGs as tracers for a first
calculation of the merger geometry, we find a projected separation of
118\arcs\ (535 kpc) and a line-of-sight velocity difference of 565
km/s. For comparison, the projected separation in the X-ray selected
Bullet cluster is 720 kpc \citep{BradacBulletLensing2006,Clowe06} and
those in \citet{MCCsampleanalysis} radio-selected sample of merging
clusters are generally closer to 1 Mpc, indicating that more time
since pericenter (TSP) has passed.  This suggests the potential of
optical selection to find systems with smaller separations hence less
TSP. Because a complete understanding of the merging process will
require snapshots of systems spanning a range of TSP, optical
selection may find its place within a range of complementary
selection methods.


\textit{Richness and related estimates.}  \citet{Rykoff2014} give the
optical richness $\lambda$ (a measure of how many galaxies are in the
cluster, within a certain luminosity range below the BCG) as
128. \cite{RMmassrichnessrelation} calibrated the relation between
weak lensing mass and $\lambda$ (including its scatter), from which we
estimate the mass of Abell 56 to be
$M_{200}=10.4 ^{+7.6}_{-4.6}\times10^{14}\ h^{-1}$
M$_\odot$. \citet{massproxies2017} implemented a system for mass
forecasting with proxies, taking into account various biases, and
found $M_{200}=11.49\pm0.89 \times10^{14}$ M$_\odot$ for this system based on its
redMapper richness. They also found $M_{500}=7.09\pm 0.77 \times10^{14}$ M$_\odot$
using $Y_{\rm 500}$, a measure of the Sunyaev-Zel'dovich effect, as a
proxy. For comparison, \citet{PSZ2} found
$M_{500}=5.62^{+0.54}_{-0.58} \times10^{14}$ M$_\odot$ from their
scaling relation based on the same $Y_{\rm 500}$ measurement.

From the scaling relations of \citet{RMscalingrelations2014} one would
expect the X-ray temperature $T_X$ to be around 7 keV with up to
40\% scatter at fixed richness. Because this is a merging cluster, the
X-ray properties may vary from the scaling relations even more than
usual.

\textit{X-ray properties from archival data.} The cluster was observed
with the \XMM\ European Photon Imaging Camera (EPIC) in 2010 (Obs.ID
0650380401, P.I. Allen). The exposure times were 7121 s, 7127 s, and
5533 s for the MOS1, MOS2, and PN instruments, respectively. As the
short exposure does not allow for a detailed analysis of the
intracluster medium (ICM) properties and the cluster morphology, we
restricted our analysis to obtaining a point-source-subtracted,
exposure-corrected image, as well as a global temperature and
luminosity for the cluster.

We performed the data reduction using the \XMM\ Science Analysis
System (\texttt{SAS}) version 19.0.0. We excluded periods of high
soft-proton background by imposing a cutoff of 0.4 (0.8) on the
soft-proton rate for the MOS (PN) detectors\footnote{We define MOS
  (PN) soft-proton events as those with energy $E{>}10$
  ($10{<}E{<}12$) keV. The higher-than-usual baseline soft-proton rate
  for this observation may result in significant residual soft-proton
  contamination even after the exclusion of flare events. This is at
  least partially mitigated by the background-subtraction strategy.},
which resulted in filtered exposure times of 6817 s, 6621 s, and 4178
s for MOS1, MOS2, and PN, respectively. Only single-to-quadruple
events from MOS and single-to-double events from PN were used in our
analysis.  Point-source detection and masking were performed by the
\texttt{cheese} routine from the \texttt{ESAS} package.  The contours
in Figure~\ref{fig-ovw} are from the 0.4-1.25 keV band after
point-source masking and exposure correction, using the procedure
described in the XMM ESAS Cookbook \citep{ESAS} and adaptively
smoothed using the \texttt{adapt} routine from \texttt{ESAS}.

In order to obtain a background-subtracted spectrum, we used the
double-subtraction method described in \cite{Arnaud2002}. We defined
the source region as a circle with a 90\arcs\ radius centered on the
cluster, whereas the background was extracted from a slightly larger
circular region away from the cluster. Blank-sky files
\citep{Carter2007} were used to mitigate the effects of the spatial
variation of background components across the detector. For a detailed
description of the method, we refer to \cite{Arnaud2002}. Using
\textsc{XSPEC} \citep{XSPEC}, we fit the spectrum to an \texttt{apec}
model, multiplied by a \texttt{phabs} model to account for galactic
absorption. We obtained a total unabsorbed luminosity in the $0.5-10.0$ keV range
of $L_X=3.8\pm0.2\times 10^{44}$ erg/s and a temperature
$T_X=5.9^{+1.1}_{-0.8}$ keV, where the uncertainties represent the
$90\%$ confidence intervals.

\section{Redshift survey and clustering kinematics}\label{sec-z}

\subsection{Redshift survey}

\textit{Observational setup.} We observed Abell 56 with the DEIMOS
multi-object spectrograph \citep{FaberDEIMOS} at the W. M. Keck
Observatory on July 1, 2022 (UT).  The DEIMOS field of view is
approximately 16$^\prime \times 4^\prime$, making it well suited to
merging clusters when the long axis is placed along the subcluster
separation vector.  We prepared two slitmasks with approximately sixty
1$^{\prime\prime}$ wide slits in each.  Galaxies were selected for
targeting based on (i) a preference for brighter targets; and (ii) a
preference for galaxies likely to be in the cluster based on
Pan-STARRS photometric redshifts \citep{PSphotoz2021}. Because the
photometric redshifts are imprecise, this approach naturally helps
probe for potential foreground/background structures that could affect
the modeling of Abell 56. Specifically, each Pan-STARRS photometric
redshift $z_{\rm PS}$ has a corresponding uncertainty
$\sigma_{\rm PS}$ such that the likelihood of the galaxy being in a
cluster at redshift $z_{\rm cl}$ is 
\begin{equation}
  \label{eq:1}
\mathcal{L}\propto\frac{1}{ \sigma_{\rm  PS}}  \exp \frac{(z_{\rm PS}-z_{\rm cl})^2}{2 \sigma_{\rm
  PS}^2}
\end{equation}
The median value of $\sigma_{\rm PS}$ was 0.16, so a broad range of
redshifts was included. We then upweighted brighter galaxies by
applying a multiplicative weight $(24-r)$, where $r$ is the apparent
$r$ magnitude, to quantify the priority of each galaxy as input to the
slitmask design software \texttt{dsimulator}; larger numbers indicate
higher priority. We manually raised the priority of a few galaxies
that potentially formed a foreground group at the north end of the
field.

We used the 1200 line mm$^{-1}$ grating, which results in a pixel
scale of 0.33 \AA\ pixel$^{-1}$ and a resolution of ${\sim}1$ \AA\ (50
km/s in the observed frame). The grating was tilted to observe
the wavelength range $\approx$ 4200--6900 \AA (the precise range
depends on the slit position), which at $z\approx0.3$ includes
spectral features from the [OII] 3727 \AA\ doublet to the magnesium
line at 5177 \AA. The total exposure time was 45 (77) minutes on the
first (second) mask, divided in three (four) exposures. The seeing was
roughly 1\arcsec, with minor variations over time.

\textit{Data reduction and redshift extraction.} We calibrated and
reduced the data to a series of 1-D spectra using PypeIt
\citep{pypeit:joss_pub,pypeit:zenodo}.  We double-checked the arc lamp
wavelength calibration against sky emission lines, and found good
agreement.

To extract redshifts from the 1-D spectra we wrote custom Python software to
emulate major elements of the approach used by the DEEP2
\citep{Deep2:2013} survey using the same instrument.  The throughput
as a function of wavelength varies from slit to slit, hindering direct
comparison to template spectra. Because throughput is generally a
slowly varying function of wavelength, the spectra are compared to
templates only after removing the slowly varying trends from each.
First, telluric absorption features are reversed using Mauna Kea
models from the PypeIt development suite. Next, we create a smooth
model or unsharp mask by convolving the 1-D spectrum with a kernel
150 \AA\ wide, which is uniform but for a 10 \AA\ diameter hole in the
center. Finally, the intensity of each pixel in the 1-D spectrum is
expressed as a fraction of the intensity in the smooth model.

The same operations are performed on redshifted versions of the galaxy
templates from the Sloan Digital Sky Survey\footnote{Available at
  \url{https://classic.sdss.org/dr7/algorithms/spectemplates/}; we
  used templates 23 through 27.}, and a $\chi^2$ value is computed for
each template-redshift combination. A user then inspects the match
between the data and the model with the global minimum $\chi^2$, or
other models at local minima, before determining whether a redshift is
secure. A secure redshift may not appear at the global minimum due to
poorly subtracted sky lines or other artifacts (e.g., a spurious
``line'' with spuriously small uncertainties may appear at the gap
between CCDs). Furthermore, some slits suffer from vignetting at the
red end, which appears as a drop in intensity too steep to be removed
by the unsharp masking process. In these cases the user specifies a
maximum wavelength to consider for template matching. A negligible
fraction of slits contained stars, so we did not include stellar
templates in the automated search; users can manually classify a
spectrum as a star without extracting a redshift.

The uncertainty in the redshift is initially computed from the
curvature of the $\chi^2$ surface about the minimum, and is typically
$\lesssim10^{-4}$ (23 km/s in the frame of the cluster). We compared
redshifts obtained by different users on different computing hardware,
operating systems, and Python installations.  We found that
user-dependent uncertainty is also $\lesssim10^{-4}$, mostly due to
specification of the maximum wavelength. We therefore add
$10^{-4}$ in quadrature to the uncertainty derived from the
curvature of the $\chi^2$ surface to derive a final uncertainty
estimate. We found 54 and 48 secure redshifts in the two masks
respectively, for a total of 102. These are listed in
Table~\ref{tab-zspec}.

\begin{deluxetable}{llll}
  \tablecaption{Galaxy redshifts}  \label{tab-zspec}
  \tablecolumns{4}
  \tablehead{\colhead{RA (deg)} & \colhead{DEC (deg)} & \colhead{z} & \colhead{uncertainty}}
  \startdata
8.410133 & -7.758144 & 0.370305 & 0.000107\\ 
8.412567 & -7.747167 & 0.370005 & 0.000102\\ 
8.411033 & -7.742000 & 0.370055 & 0.000103\\ 
8.412454 & -7.750711 & 0.411192 & 0.000182\\ 
8.414025 & -7.767139 & 0.361506 & 0.000100\\ 
8.418721 & -7.738369 & 0.124575 & 0.000103\\ 
8.419221 & -7.752617 & 0.309164 & 0.000132\\ 
8.424929 & -7.768175 & 0.550809 & 0.000390\\ 
8.439262 & -7.728628 & 0.342070 & 0.000149\\ 
8.449196 & -7.770197 & 0.300759 & 0.000103\\ 
8.451654 & -7.772564 & 0.301399 & 0.000100\\ 
8.483425 & -7.776081 & 0.305369 & 0.000106\\ 
8.430800 & -7.878178 & 0.285649 & 0.000108\\ 
8.433154 & -7.865233 & 0.302410 & 0.000111\\ 
8.436871 & -7.800781 & 0.303911 & 0.000101\\ 
8.437129 & -7.820883 & 0.307063 & 0.000103\\ 
8.439008 & -7.874475 & 0.310816 & 0.000104\\ 
8.442646 & -7.833144 & 0.295355 & 0.000102\\ 
8.442646 & -7.833144 & 0.304912 & 0.000273\\ 
8.442892 & -7.838442 & 0.296056 & 0.000101\\ 
8.443604 & -7.795317 & 0.368704 & 0.000100\\ 
8.446533 & -7.918775 & 0.510949 & 0.000554\\ 
8.446804 & -7.847947 & 0.302537 & 0.000120\\ 
8.446900 & -7.865711 & 0.292854 & 0.000103\\ 
8.448988 & -7.939147 & 0.301960 & 0.000111\\ 
8.451133 & -7.910875 & 0.305312 & 0.000107\\ 
8.448746 & -7.942206 & 0.764905 & 0.000207\\ 
8.452063 & -7.952244 & 0.364001 & 0.000107\\ 
8.455208 & -7.976750 & 0.306263 & 0.000109\\ 
8.454325 & -7.842231 & 0.291603 & 0.000101\\ 
8.457279 & -7.888228 & 0.309365 & 0.000103\\ 
8.457279 & -7.888228 & 0.309039 & 0.000103\\ 
8.457421 & -7.879375 & 0.300609 & 0.000104\\ 
8.460504 & -7.866011 & 0.304762 & 0.000103\\ 
8.460571 & -7.821856 & 0.312767 & 0.000104\\ 
8.462892 & -7.842867 & 0.299458 & 0.000100\\ 
8.463775 & -7.837614 & 0.304938 & 0.000101\\ 
8.463446 & -7.852147 & 0.312048 & 0.000101\\ 
8.465396 & -7.800806 & 0.209615 & 0.000103\\ 
8.466696 & -7.805417 & 0.166386 & 0.000104\\ 
8.466279 & -7.910831 & 0.300395 & 0.000119\\ 
8.466696 & -7.805417 & 0.303377 & 0.000103\\ 
8.468108 & -7.882522 & 0.301556 & 0.000102\\ 
8.469058 & -7.866528 & 0.308831 & 0.000108\\ 
8.469108 & -7.846475 & 0.300255 & 0.000108\\ 
8.470600 & -7.894619 & 0.300445 & 0.000115\\ 
8.469867 & -7.922953 & 0.299295 & 0.000100\\ 
8.472408 & -7.810456 & 0.368080 & 0.000101\\ 
8.479588 & -7.966142 & 0.301866 & 0.000104\\ 
8.487733 & -7.926608 & 0.301159 & 0.000101\\ 
8.491658 & -7.896106 & 0.297784 & 0.000100\\ 
8.494708 & -7.913458 & 0.307990 & 0.000874\\ 
8.497242 & -7.883722 & 0.298694 & 0.000107\\ 
8.512950 & -7.938439 & 0.533748 & 0.000110\\ 
8.440079 & -7.972850 & 0.304762 & 0.000106\\ 
8.438658 & -7.847369 & 0.302260 & 0.000147\\ 
8.440954 & -7.950100 & 0.761086 & 0.000358\\ 
8.441692 & -7.832436 & 0.295155 & 0.000102\\ 
8.441692 & -7.832436 & 0.304628 & 0.000120\\ 
8.444413 & -7.918158 & 0.301403 & 0.000102\\ 
8.444925 & -7.742297 & 0.298120 & 0.000119\\ 
8.445521 & -7.845039 & 0.643521 & 0.000100\\ 
8.446550 & -7.870319 & 0.303561 & 0.000101\\ 
8.443504 & -7.981761 & 0.390419 & 0.000106\\ 
8.449642 & -7.842747 & 0.371022 & 0.000101\\ 
8.448617 & -7.785400 & 0.306663 & 0.000103\\ 
8.453133 & -7.920878 & 0.293771 & 0.000101\\ 
8.455708 & -7.876181 & 0.412010 & 0.000107\\ 
8.458538 & -7.838819 & 0.316319 & 0.000101\\ 
8.458929 & -7.864583 & 0.303911 & 0.000103\\ 
8.460292 & -7.783261 & 0.263697 & 0.000101\\ 
8.459621 & -7.840900 & 0.295088 & 0.000100\\ 
8.467704 & -7.995767 & 0.449141 & 0.000103\\ 
8.463604 & -7.836289 & 0.304228 & 0.000104\\ 
8.464392 & -7.744386 & 0.127610 & 0.000100\\ 
8.464413 & -7.885931 & 0.298040 & 0.000102\\ 
8.466088 & -7.898686 & 0.299508 & 0.000104\\ 
8.465567 & -7.866972 & 0.302907 & 0.000101\\ 
8.467179 & -7.830192 & 0.302136 & 0.000114\\ 
8.470292 & -7.732906 & 0.457597 & 0.000110\\ 
8.467813 & -7.806775 & 0.166486 & 0.000103\\ 
8.471492 & -7.765292 & 0.458958 & 0.000101\\ 
8.471050 & -7.773569 & 0.276543 & 0.000101\\ 
8.474892 & -7.873158 & 0.307967 & 0.000100\\ 
8.476692 & -7.907672 & 0.368621 & 0.000100\\ 
8.477433 & -7.955953 & 0.369154 & 0.000103\\ 
8.480146 & -7.883764 & 0.296739 & 0.000107\\ 
8.478713 & -7.778783 & 0.305796 & 0.000104\\ 
8.481079 & -7.796372 & 0.370439 & 0.000102\\ 
8.481037 & -7.852150 & 0.303194 & 0.000102\\ 
8.479967 & -7.927103 & 0.292623 & 0.000104\\ 
8.484304 & -7.771003 & 0.469204 & 0.000102\\ 
8.485038 & -7.902189 & 0.412787 & 0.000105\\ 
8.486742 & -7.826083 & 0.369608 & 0.000111\\ 
8.487963 & -7.821725 & 0.367643 & 0.000108\\ 
8.487963 & -7.821725 & 0.367550 & 0.000115\\ 
8.488279 & -7.849608 & 0.304134 & 0.000101\\ 
8.490088 & -7.816639 & 0.302260 & 0.000101\\ 
8.494679 & -7.892922 & 0.144521 & 0.000101\\ 
8.497133 & -7.756675 & 0.290352 & 0.000108\\ 
8.500467 & -7.915733 & 0.458747 & 0.000103\\ 
8.503554 & -7.811536 & 0.371723 & 0.000105\\ 

    \enddata
\end{deluxetable}

\begin{figure}
\centerline{\includegraphics[width=\columnwidth]{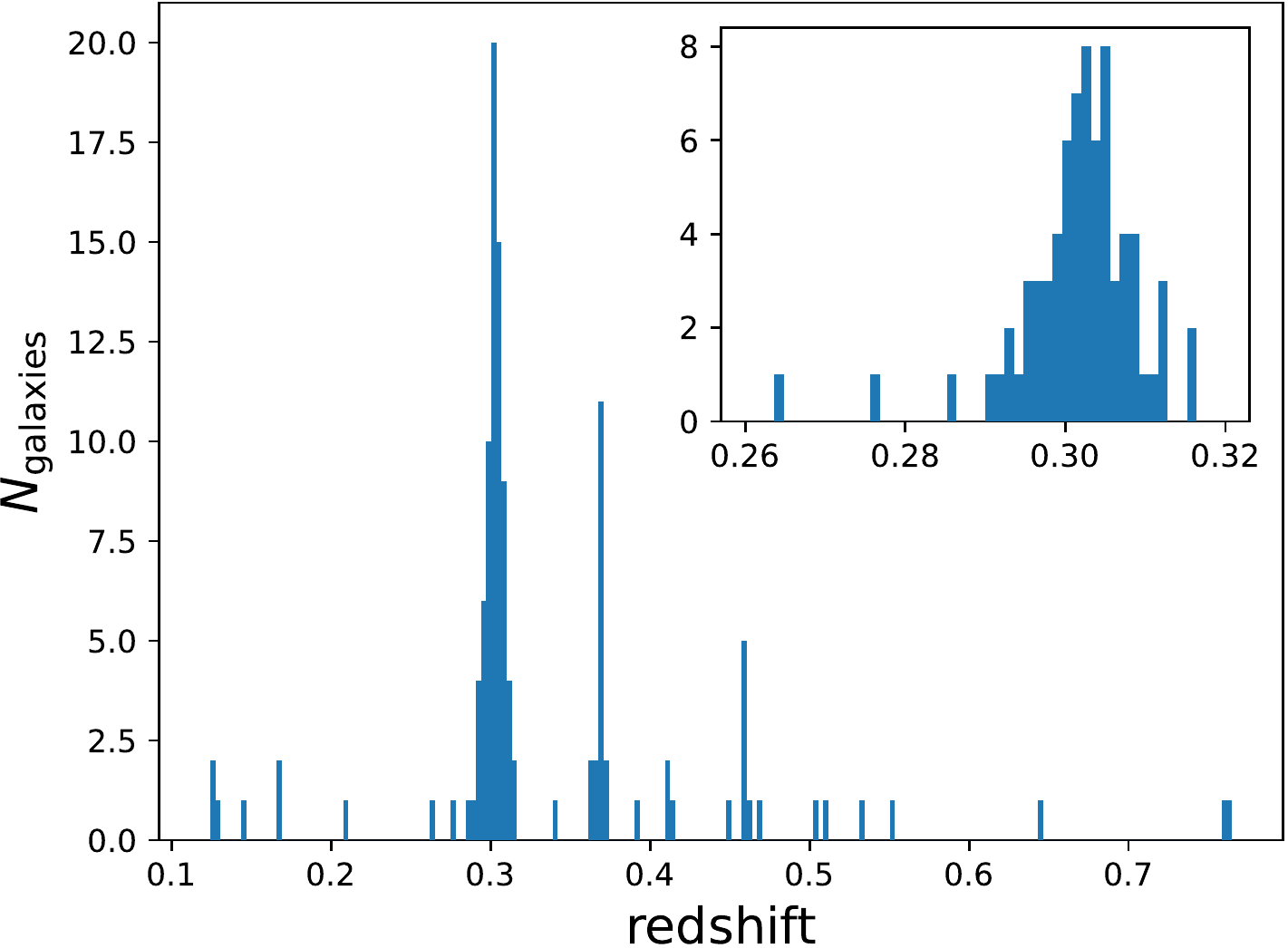}}
\caption{Redshift histogram, with inset showing the redshift interval
  around Abell 56.}
\label{fig-zhist}
\end{figure}

\textit{Comparison to archival redshifts.} We searched NED for
archival spectroscopic redshifts within a radius of 5\arcm, and found
16 galaxies, largely from BOSS \citep{BOSSoverview2013}. Of these,
five were galaxies that we had targeted.  The mean redshift difference
between independent measurements of the same target is $9$ km/s, with
an rms scatter of $9$ km/s.  We then removed the duplicates and merged
the catalogs from NED and from our two masks to produce a final
catalog of 113 galaxies. Figure~\ref{fig-zhist} shows a histogram of
these redshifts.

\subsection{Subclustering and kinematics}\label{ssec-kinematics}

\begin{figure}
\centerline{\includegraphics[width=\columnwidth]{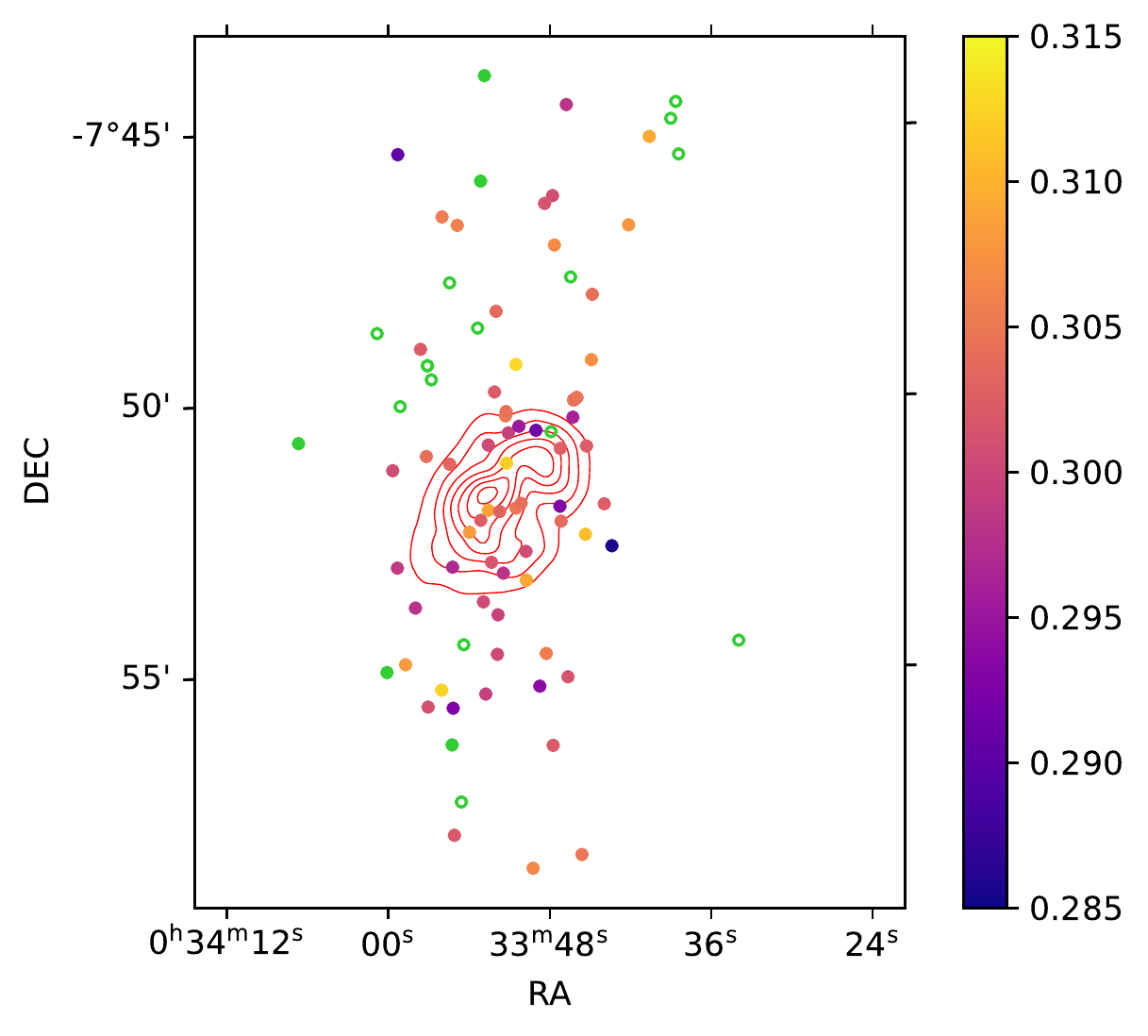}}
\caption{Redshift map. Abell 56 galaxies are coded with a continuous
  color map, while galaxies in the putative background cluster at
  $z=0.37$ (0.46) are shown as open (closed) green circles. \XMM\
  contours are shown in red.}
\label{fig-zmap}
\end{figure}

\textit{Non-Abell 56 structures.} Figure~\ref{fig-zhist} reveals a
potential background cluster at $z=0.37$, and possibly another at
$z=0.46$. To assess how strongly clustered these galaxies are on the
sky, Figure~\ref{fig-zmap} shows a sky map color-coded by
redshift. Galaxies in the putative cluster at $z=0.37$ (0.46) are
shown as open (closed) green circles, while galaxies near $z=0.30$
(i.e., associated with Abell 56) are shown with a continuous color map
that contains no green. Neither set of background galaxies shows signs
of clustering in space. Furthermore, we estimate the velocity
dispersion of each set using the biweight estimator \citep{Beers1990}
and find only $279\pm 45$ ($105 \pm 48$) for the structure at $z=0.37$
(0.46). Uncertainties on biweight estimators are obtained by the
jackknife method throughout this paper. These velocity dispersions are
far less than the velocity dispersion of Abell 56 (below), suggesting
that they are an order of magnitude less massive and unlikely to
substantially contaminate the weak-lensing and X-ray maps presented
below.

\textit{Abell 56.} Of 67 galaxies in the $0.285\le z\le 0.314$ window,
the biweight estimate for the systemic redshift is
$0.30256\pm 0.00058$. At this redshift, the physical scale is 4.521
kpc/arcsec given our adopted cosmological model
\citep{Wright2006CosmologyCalculator}.  The redshift distribution is
compatibile with a single Gaussian, according to a Kolmogorov-Smirnov
test.  This is consistent with the low line-of-sight velocity
component \dv\ suggested by the archival redshifts of the north and
south BCGs.  The biweight estimate of velocity dispersion is
$1264\pm145$ km/s---rather large, but likely to be inflated by merger
activity as noted below.

\begin{figure}
\centerline{\includegraphics[width=\columnwidth]{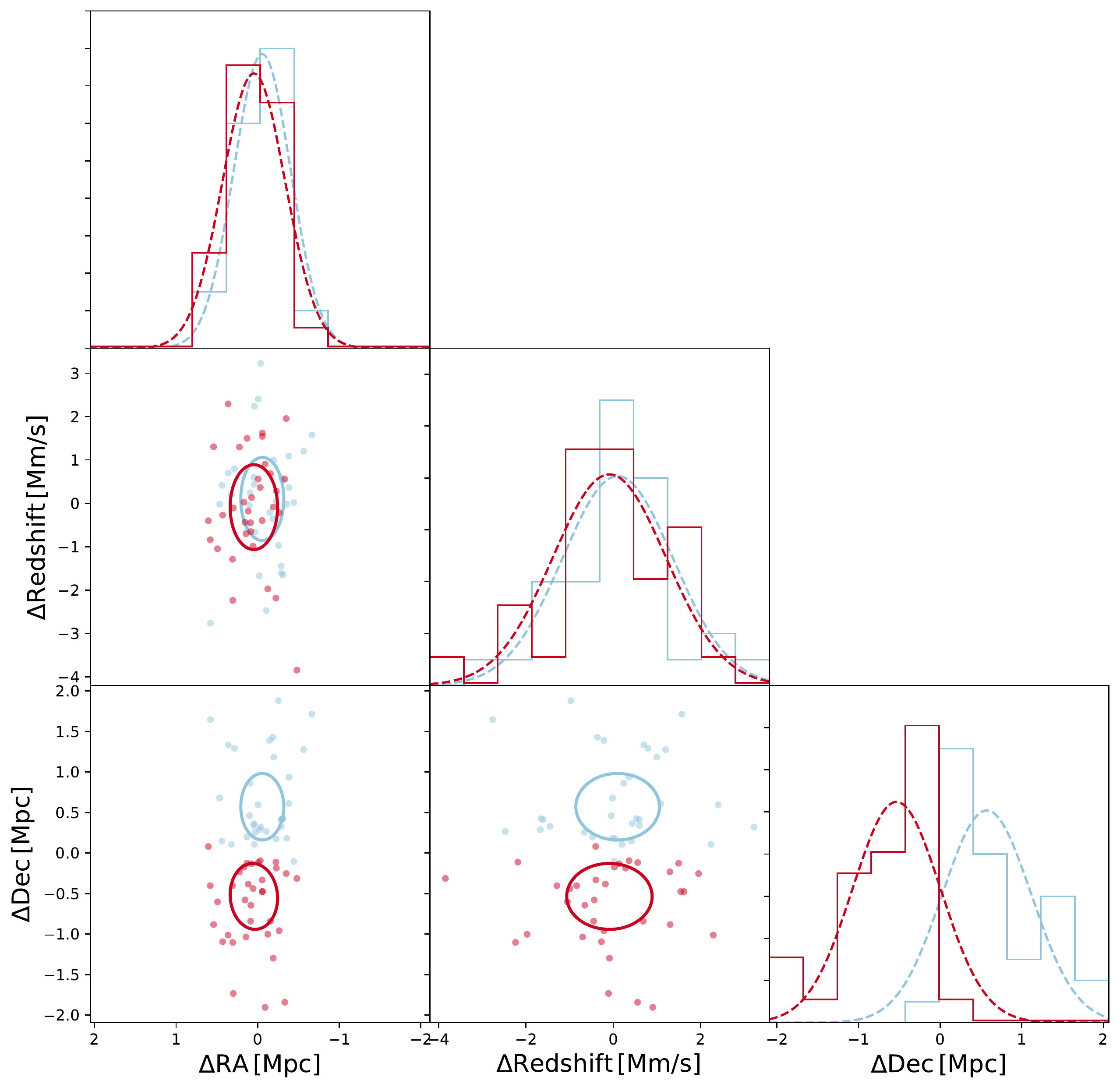}}
\caption{Corner plot showing distribution of subcluster members in RA,
  DEC, and velocity space relative to the overall mean. Members of the
  north (south) subcluster are shown in blue (red).}
\label{fig-corner}
\end{figure}

We use the \texttt{mc3gmm} code \citep{MCCsampleanalysis} to assign
galaxy membership to subclusters. This code models the
  distribution of galaxies in (RA, Dec, z) space as a mixture of $N$
  elliptical Gaussian profiles (i.e., subclusters), with physically
  motivated priors on subcluster variance in each dimension as well as
  covariance (i.e., ellipticity and rotation).  $N$ is determined by
  the user; we set $N{=}2$ based on the optical imaging and further
  supported by the lensing data presented in
  \S\ref{sec-wl}.\footnote{\citet{MCCsampleanalysis} varied
      $N$ and for each merging system found the value of $N$ that best
      satisfied the Bayesian Information Criterion (BIC), thus deriving $N$
      from the spectroscopic data alone. Here, the subcluster
      separation is smaller than typically seen in
      \citet{MCCsampleanalysis}, and the spectroscopic data points are
      fewer, making it more difficult to meet BIC criteria
      for $N{>}1$ based on the
      spectroscopy alone.}  The user sets nonoverlapping bounds for
  the central (RA, Dec, z) of each subcluster to avoid degeneracies;
  \texttt{mc3gmm} maximizes the likelihood by adjusting the parameters
  within those bounds.  We run \texttt{mc3gmm} on the galaxies in the
  redshift window $0.285\le z\le 0.314$ and the result is shown in
Figure~\ref{fig-corner}. The velocities of the subclusters are nearly
identical, suggesting that the relative motion of the subclusters is
in a direction close to the plane of the sky. The biweight estimate
for the systemic redshift of the 33 (34) north (south) members is
$0.30298\pm0.00099$ ($0.30231\pm 0.00071$). This yields
$\dv=153\pm281$ km/s.

The biweight velocity dispersion is $1283\pm236$ km/s for the north
subcluster, and $1251\pm191$ for the south.  Simulations of merging
clusters \citep[e.g.,][]{Pinkney96,Takizawa10} show that a pericenter
passage in the plane of the sky boosts the observed velocity
dispersion by a factor of ${\approx}1.5$ for hundreds of Myr
afterward. Hence, one should not interpret these large velocity
dispersions as indicative of extremely massive clusters.

\section{Weak lensing analysis}\label{sec-wl}

We perform a weak-lensing analysis on the HST F814W imaging. Galaxies
are detected in the F814W image with SExtractor \citep{1996bertin}.
For each galaxy, PSF models are generated following the method of
\cite{2007jee} by utilizing their publicly available PSF catalog. Each
galaxy is fit with a PSF-convolved Gaussian distribution and the
complex ellipticities are recorded \citep[our ACS weak-lensing
pipeline is outlined in][]{2017finner, 2021finner,
  2023finner}. Objects with ellipticity greater than 0.8, ellipticity
uncertainty greater than 0.3, and intrinsic size (pre-psf) less than
0.5 pixels are removed to prevent spurious sources such as diffraction
spikes around bright stars and poorly fit objects from entering the
source catalog.

The next step is to eliminate as many foreground and cluster galaxies
as possible, while still retaining a sizeable sample of background
sources. With only single-band imaging available, we select galaxies
with F814W AB magnitudes fainter than 24. We apply this magnitude cut
to the GOODS-S photometric redshift catalog \citep{GOODSzphot2013} and
find that the contamination by foreground galaxies is expected to be
$\sim2\%$. Cluster galaxies may contribute additionally to the
contamination. As their contamination should be radially dependent, we
test the radial dependence of the source density. We find it to be
flat, which suggests cluster galaxies are not significantly
contaminating our source catalog. The final source catalog contains
$\sim43$ galaxies arcmin$^{-2}$. The source catalog is then provided
to the \texttt{FIATMAP} code \citep{2006wittmanDLS} to create a
surface mass density map. \texttt{FIATMAP} convolves the
  observed shear field with a kernel of the form
$$ r^{-2} (1-\exp({-r^2 \over 2r_i^2})) \exp({-r^2 \over 2r_o^2})
\eqno{(1)}$$ where $r_i$ and $r_o$ are inner and outer cutoffs,
respectively. The inner cutoff is necessary to prevent amplification
of shape noise in sources at small $r$, and was set to 50 arcsec. The
outer cutoff suppresses noise that may come from unrelated structures
along the line of sight at large projected separations, and is of
limited value in a small field; we set it to 100 arcsec, which is
comparable to the radius of the field.  The results were pixelized
onto a map with 1.5 arcsec pixels.  In addition to this fiducial map,
a family of viable reconstructions can be made by bootstrap resampling
the shear catalog (see below). Figure~\ref{fig-wlmap} shows the fiducial map as a set of
contours overlaid on a Pan-STARRS multiband image\footnote{Retrieved
  from \url{http://ps1images.stsci.edu/cgi-bin/ps1cutouts}.}
\citep{PSimages2020}. Two weak lensing peaks are evident, associated
with (albeit slightly offset from) each galaxy subcluster, with the
X-ray peak in between. This confirms the basic merger scenario
developed above.

\begin{figure}
\centerline{\includegraphics[width=\columnwidth]{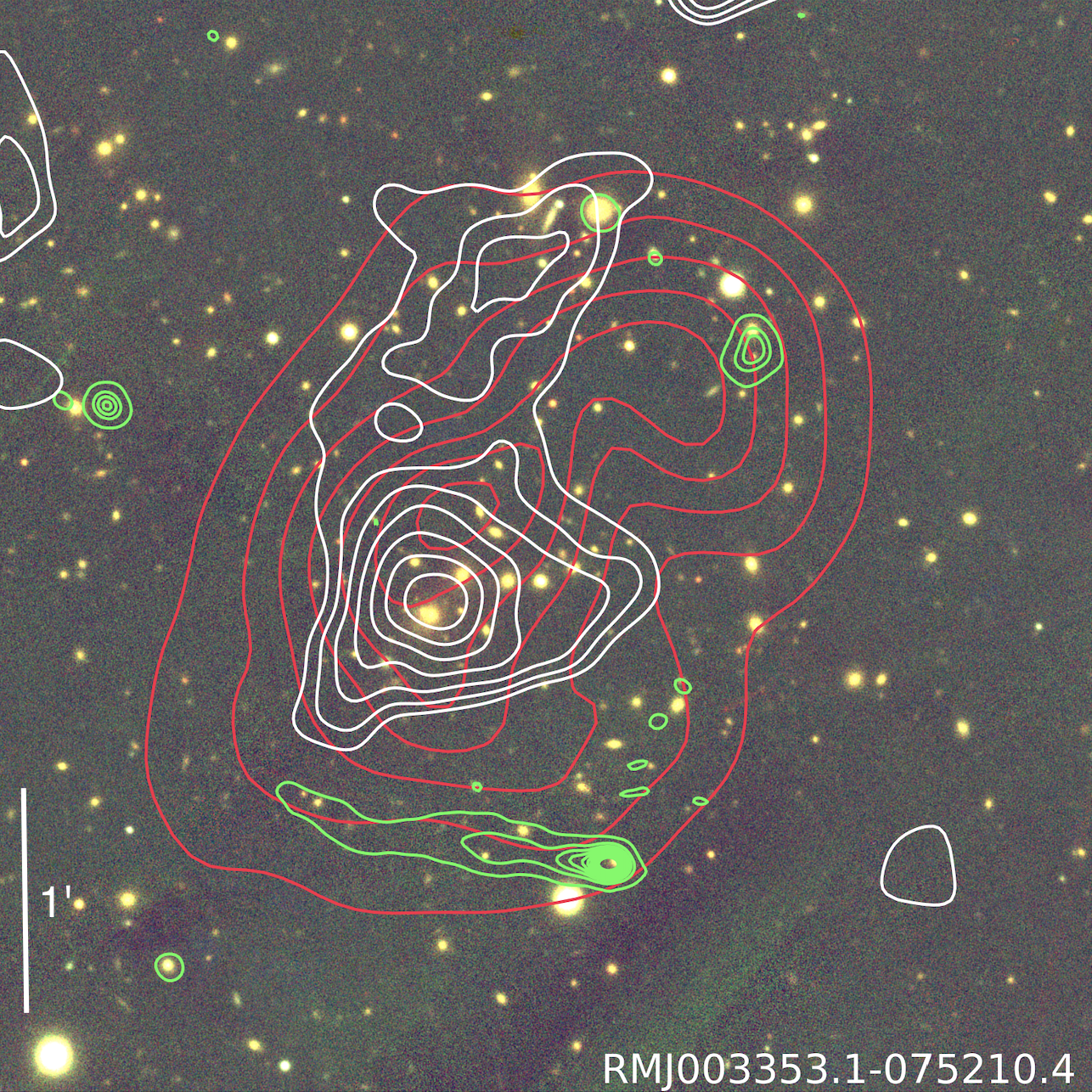}}
\caption{Surface mass density contours from weak lensing (white)
  overlaid on a Pan-STARRS multiband image and red \XMM\ surface brightness
  contours. The small closed contour between the subclusters is a
  trough. Green GMRT 650 MHz contours
  (\S\ref{sec-GMRT}) start at 70 $\mu$Jy/beam with
  increments of 680 $\mu$Jy/beam and a 4\arcs\ beam.}
\label{fig-wlmap}
\end{figure}

We estimate the mass of each subcluster by fitting a two-halo NFW
model with a fixed mass-concentration relation from
\cite{2019diemer}.  To achieve the best-fit model, the shear of the two-halo
model is derived at the position of each background galaxy and the
chi-square is minimized.  The effective distance ratio of the sources is set
by the effective distance ratio of GOODS-S sources fainter than 24th
magnitude.  The best-fit two-halo model has a mass of
$M_{200}=4.5\pm0.8\times10^{14}$ M$_\odot$ and
$M_{200}=2.8\pm0.7\times10^{14}$ M$_\odot$ for the south and north
subclusters, respectively. We allow the centroid of each halo to be
fit and they converge to the projected mass distribution peaks. On the
other hand, if we fix the halo centroids to the BCGs, we find the
south and north subcluster masses decrease by $10\%$ and $60\%$,
respectively.  To test the dependence of the mass estimate on our
choice of magnitude cut, we vary the magnitude constraint on the
background catalog from 22nd to 25th magnitude and find that the mass
estimates decrease for brighter magnitude cuts but within the mass
uncertainty. To estimate the total mass of the cluster, we simulate
two NFW halos at the projected separation of the two mass
peaks. Integrating the model from the center of mass to $R_{200}$, we
estimate the total mass of the cluster to be
$M_{200}=9.7\pm2.0\times10^{14}$ M$_\odot$
($M_{500}=7.1\pm1.6\times10^{14}$ M$_\odot$).

To further quantify the detection significance, we bootstrap resampled
the source catalog to generate 1000 realizations of the mass map.
As expected, the mean map yielded by these resamplings matches
  the fiducial map yielded by the original catalog.  At any given sky
  position, we can measure the rms variation of surface mass density
  across the map realizations to obtain a noise map. The ratio of the
  fiducial map to this noise map is then a significance map.  The peak
  of the southern (northern) subcluster is detected at a significance
  of $6.3$ ($5.5$).

The projected separation between the mass peaks, $\dproj=438$ kpc, is
important for the dynamical modeling in \S\ref{sec-analogs}.  To
estimate the uncertainty on the peak locations, the peak from each of
the 1000 realizations was recorded. The 1000 peaks were then passed to
a $k$-means algorithm with the number of distributions fixed to
two. The $k$-means algorithm iteratively calculates the centroid of
the peaks and assigns peaks to each centroid until the centroid
converges. This procedure yields two distributions of mass-peak
locations, which are then processed with a kernel density estimator to
find the $1\sigma$ and $2\sigma$ uncertainties. We find that the
southern mass peak is consistent with its BCG at the $1\sigma$
level. In contrast, the northern mass peak is offset $19.2\pm4.9$ arcsec
($87\pm22$ kpc) to the south of the northern BCG. We address this offset
further in \S\ref{sec-DM}. Our immediate goal here is to define a 68\%
confidence interval on the projected separation between mass peaks,
which we find to be $96.9\pm45.6$ arcsec ($438\pm206$ kpc).


To further check the halo position uncertainties, we consider again
the two-halo fit.  As a model-driven procedure, this should be more
robust against edge effects than the mapping procedure, which
convolves the observed shear field.  Nevertheless, as noted above, the
halo center parameters converge to the projected mass distribution
peaks.  The positional uncertainties from the two-halo fit are smaller
than those from the resampled mapping method. Hence our adoption of
the values from the latter method is the more cautious approach.

\section{Simulated analogs and dynamical parameters}\label{sec-analogs}
  
We find analog systems in the Big Multidark Planck (BigMDPL)
 Simulation \citep{BigMDPL2016} using the method of
 \citet{WCN18analogviewingangle} and \citet{Wittman19analogs}.
The observables used to constrain the likelihood of any given
analog and viewing angle are:
\begin{itemize}
\item the projected separation between mass peaks \dproj, for which we use
  $438\pm206$ kpc from \S\ref{sec-wl}.
\item the line-of-sight relative velocity \dv, for which we use
  $153\pm281$ km/s from \S\ref{sec-z}.
\item the subcluster masses, for which we use
$M_{200}=4.5\pm0.8\times10^{14}$ M$_\odot$ and
$M_{200}=2.8\pm0.7\times10^{14}$ M$_\odot$ for the south and north
subclusters, respectively, from \S\ref{sec-wl}.  Note that dynamical
timescales and velocities depend only weakly on the masses.
\end{itemize}

Table~\ref{tab-dynprop} lists the resulting highest probability
density confidence intervals for time since pericenter (TSP),
pericenter speed $v_{\rm max}$, viewing angle $\theta$ (defined as the
angle between the subcluster separation vector and the line of sight,
i.e. 90$^\circ$ when the separation vector is in the plane of the
sky), and the angle $\varphi$ between the current separation and
velocity vectors. $\varphi$ is potentially an indicator of how head-on
the trajectory is, as well as of merger phase (surpassing 90$^\circ$
at apocenter).  The likelihood ratio of analogs in the outbound
vs. returning phase is 19:1. Table~\ref{tab-dynprop} also lists the
confidence intervals for the dynamical parameters when the analysis is
restricted to the outbound scenario. These particular parameters are
not sensitive to the current merger phase.

\begin{table}
  \centering  
  \caption{Dynamical parameters from analogs}
  \begin{tabular}{ccccc}
   Scenario &  TSP (Myr) & $v_{\rm max}$ (km/s) & $\theta$ (deg) &
                                                                   $\varphi$
                                                                   (deg)\\
    \hline
    \multicolumn{5}{c}{68\% CI}\\ \hline
    All & 60-271 &1960-2274 &68-90 &6-33\\
    Outbound & 90-291 &1952-2282 &68-90 &0-26\\
    \multicolumn{5}{c}{95\% CI}\\ \hline
    All & 0-451 &1729-2510 &42-90 &0-86\\
    Outbound& 0-366 &1681-2489 &44-90 &1-67\\
  \end{tabular}
  \label{tab-dynprop}
\end{table}

\section{Radio observations and results}\label{sec-GMRT}

Pericenter speeds in cluster mergers are typically greater than the
sound speed in the gaseous ICM, so each subcluster launches a shock in
the ICM of the other subcluster \citep{shocks2018}.  In
  hydrodynamic simulations of the Bullet \citep{Springel2007} the
  shock begins at pericenter speed and loses very little speed over
  time, while the corresponding subcluster falls behind due to the
  gravity of the other subcluster.  Our Abell 56 analogs do not
  include gas, but we use the pericenter speed, gravitational
  subcluster slowing, and analog time of observation to predict the
  separation between a subcluster and a hypothetical constant-velocity
  shock. We find ${\sim}200$ kpc separation in the outbound phase. The
  analogs indicate that an additional ${\approx}1.2$ Gyr passes before
  the subclusters return to the same projected separation en route to
  a second pericenter. In this time, a hypothetical constant-velocity
  shock would have proceeded over 2 Mpc further out. Therefore,
  observing the shock location could further disambiguate between
  outbound and returning scenarios. This toy model glosses over the
  complexities of ICM properties affecting the shock speed, but the
  timescale of the returning scenario is so long that the subcluster-shock
  separation remains ${>}1$ Mpc even with
  factor-of-two variations in shock speed, or complete stalling of the
  shock after ${\sim}500$ Myr.


Shocks are often detected as discontinuities in the X-ray surface
brightness, but in this case the archival X-ray data are too shallow
to support such a detection. Shocks may also inject sufficient
non-thermal energy into charged particle motion that electrons emit
synchrotron radiation, detectable as an extended radio source known as
a radio relic \citep{ReinoutRadioReview19}. Archival 150 MHz data from
the TIFR GMRT Sky Survey (TGSS) Alternative Data Release
\citep{IntemaADR} show extended emission 270 kpc south of the southern
BCG. Due to the large synthesized beam size (25\arcs) and an
accompanying point source, it is difficult to further characterize
this emission using the TGSS data alone. \citet{Cuciti21} observed the
cluster at 1.5 GHz and $\approx12\arcs$ beam using the Jansky Very
Large Array (JVLA).  The source in question appears at the southern
edge of their Figure A.1.  However, they classified this cluster as
having no extended emission, presumably because they pointed at the
original Abell coordinates, about $7^\prime$ north of the source in
question, and because they were primarily searching for radio halos
rather than relic candidates. We also checked the VLASS \citep{VLASS2020}
and GLEAM \citep{GLEAM2015} surveys, and found no evidence of a halo or relic. 

We were granted 15 hours on the upgraded GMRT \citep[uGMRT,][]{uGMRT}
for Band 4 (550-900 MHz) observations of Abell 56 (proposal code
42\_069) with much smaller synthesized beam size (4\arcs). Observations
were taken on 20 June 2022 and 24 June 2022. We used the SPAM pipeline
\citep{IntemaSPAM} to calibrate the visibilities, and used
\texttt{wsclean} \citep{wsclean2014,wsclean2017} to create an image.
The source 270 kpc south of the southern BCG extends for $\approx420$
kpc (93\arcs) in the east-west direction and is barely resolved in
the north-south direction. Its contours are overlaid in green on the
Pan-STARRS image in Figure~\ref{fig-wlmap}.  This makes it clear that
the bright point source at the western end of the radio emission is
coincident with a galaxy; our redshift survey confirms that this
galaxy is in the cluster.

The most likely explanation for most of this emission is an AGN tail.
Given the orientation of this feature which matches that expected of a
merger shock, it is worth considering that AGN tails play a role in
the formation of some relics by providing seed electrons that are
re-accelerated by the passage of a shock
\citep[e.g.,][]{vanWeeren2017Abell3411}.  In such cases there is
spectral aging across the narrow axis of the tail in addition to the
expected aging from head to tail.  Exploring this possibility would
require high angular resolution spectral maps.  Finally, we note that
there is no evidence of a relic much further south as expected in the
returning scenario, nor of a relic on the north side of the north
subclusters.

\section{Dark matter cross section estimate}\label{sec-DM}

\citet{Spergel00} first suggested that dark matter (DM) particles may
scatter off each other in a process distinct from the interactions
with standard model particles that are probed by direct detection
experiments.  The cross section for such scattering is usually quoted
in terms of $\sDM$, the cross section per unit mass, because
the mass of the DM particle is unknown. \citet{Markevitch04} laid out
multiple physical arguments for inferring this parameter, at least at
a back-of-the-envelope level, from merging cluster observations.
Simulations \citep[e.g.,][]{Randall2008,Robertson17Bullet} are required
to properly interpret such observations. However, as a first estimate
to motivate deeper observations and perhaps simulations of Abell 56,
we present an initial back-of-the-envelope estimate.

One physical argument is that momentum exchange will slow the DM halos
relative to the galaxies, resulting in a DM-galaxy offset.
\citet{Markevitch04} developed an argument based on finding no
significant offset: requiring that the scattering depth be ${<1}$
leads to an upper limit on $\sDM$. In this case, there is a
significant offset in the north, so we turn to the method of
\citet{Harvey14} and \citet{Harvey15}, which uses the ratio of
gas-galaxy and DM-galaxy offsets. This method relies on an analogy
between DM and the much more interactive gas, so it has some
limitations, but it also reduces some sources of observational
uncertainty. Foremost, it eliminates the assumption that the surface
mass density relevant to DM scattering---the volume density integrated
along the merger axis---equals the surface mass density we can
measure, which is nearly {\it perpendicular} to the merger axis. In
fact clusters are triaxial \citep{Harvey2021clustershape} and align to
some extent with their neighboring clusters
\citep{Joachimi-alignment-review2015}, hence one may expect greater
column density along the merger axis. To the extent this pattern is
echoed by the gas, the gas analogy may reduce this systematic
error. Second, the gas analogy eliminates any uncertainty due to
viewing angle, as that angle applies equally to the gas-galaxy and
DM-galaxy separations.

The chief limitation of the gas analogy is that it breaks down over
time. SIDM simulations show that, given enough time, the galaxies
within each subcluster fall back to their associated DM---and continue
oscillating \citep{Kim17}. Around the time of apocenter between
subclusters, the DM-galaxy offset in each subcluster has a sign
opposite that predicted by the gas analogy. Hence, the gas analogy
should not be applied if the system is observed long after
pericenter. The analogs indicate that Abell 56 is observed much closer
to pericenter than apocenter, so the gas analogy is appropriate here
for a first estimate. 

In the southern subcluster, the DM-BCG separation\footnote{All
  separations in this paragraph are quoted after projecting them onto
  the merger axis, but we note that the components perpendicular to
  the merger axis are generally negligible.} is $7\pm16$ kpc and the
gas-BCG separation is $111\pm38$ kpc, yielding $\sDM=0.35\pm1.03$
cm$^2$/g, consistent with zero. In the northern subcluster, the DM-BCG
separation is $87\pm22$ kpc, while the gas-BCG separation is unclear
because it is difficult to identify a gas peak specifically associated
with the northern subcluster.  To be conservative we use the offset to
the main gas peak, $424\pm38$ kpc.  This yields $\sDM
=1.43\pm0.61$ cm$^2$/g. Multiplying the two likelihoods yields
$\sDM =1.10\pm0.64$ cm$^2$/g.

We performed a few checks on the statistical significance of the
offset in the north. None of the 1000 bootstrap realizations of the
convergence map in \S\ref{sec-wl} placed the overall mass peak as far
north as the northern BCG, and only three of them placed a local mass
peak (defined as a peak in the northern half of the field) that far
north.

We emphasize the tentative nature of the dark matter constraint. More work will
be needed to understand why the northern subcluster has a significant
DM-BCG offset while the south does not. Ground-based weak lensing, or
more space-based pointings, may be helpful to reduce any systematic
uncertainties related to the relative small footprint of the ACS
data. Deeper imaging may reveal strongly lensed sources that could
lead to more precise mass models. X-ray or radio confirmation of a
shock position could further build confidence in the merger
scenario. Even without detection of a shock, deeper data on the
overall X-ray morphology combined with hydrodynamical simulations
would greatly advance understanding of this merger.

\section{Summary and discussion}\label{sec-discussion}

We have presented a new binary, dissociative merging galaxy cluster
discovered by cross-referencing archival X-ray data with locations of
bimodal \RM\ clusters. The selection technique has the potential to be
applied more widely, as optical surveys continue to cover more area
more deeply than ever before. In particular, the southern sky may
provide new targets via the 5000 deg$^2$ Dark Energy Survey
\citep[DES;][]{DESDR1} and eventually the deeper 20,000 deg$^2$ Legacy
Survey of Space and Time \citep[LSST;][]{Scibook}.  Finding the rare
merger through pointed X-ray followup of selected candidates will
require very careful selection.  The forthcoming eROSITA X-ray survey
could enable more of a cross-correlation approach where candidates are
selected based on joint optical and X-ray properties.

This particular cluster promises to be useful for constraints on \sDM,
given that its merger axis is close to the plane of the sky and its
trajectory was sufficiently head-on to provide a substantial
separation between the gas peak and the main BCG. The lensing map
presented here is based on a single orbit of ACS time, and should be
supplemented with deeper and wider data to better understand why there
is a significant offset in the north but not in the south.
Hydrodynamical simulations could shed light on whether this could
happen in a Cold Dark Matter (CDM) scenario, perhaps with projection
effects or other complications not identified here.  Such simulations
should also be compared to deeper X-ray maps to confirm that we
understand the merger scenario.

To place this system in context with other merging clusters
  with the potential to probe \sDM, we refer to Table 1 of
  \citet{Wittman18SIDM}, which ranked the importance of various
  subclusters used in their ensemble analysis and that of
  \citet{Harvey15}.  In \citet{Harvey15}, the measurement
  uncertainties on the ``star-gas'' separation $\delta_{\rm SG}$ and the
  ``star-interacting DM'' separation $\delta_{\rm SI}$ were assumed to be
  the same for all subclusters in the ensemble. \citet{Wittman18SIDM}
  noted that this resulted in a particularly simple analytic
  expression for the (unnormalized) inverse-variance weight of a given
  subcluster in an ensemble:
  $ {\delta_{\rm SG}^2\over 1 + \delta_{\rm SI}^2/ \delta_{\rm SG}^2}$.  By
  glossing over the measurement uncertainties in any given
  observation, this quantifies the importance of a subcluster in a
  hypothetical ensemble where all subclusters are equally well
  observed.  After normalizing this weight in the same way as did
  \citet{Wittman18SIDM} for their Table 1, we find that the southern
  subcluster of Abell 56 would appear in eighth place on the list of
  usable subclusters (additional subclusters with formally greater
  weight were marked as unusable in that table due to various
  complications).  The northern subcluster of Abell 56 is difficult to
  place on this table because only an upper limit, not a measurement,
  is available for $\delta_{\rm SG}$. More X-ray data will be needed to
  determine the constraining potential of this substructure.

\acknowledgments
RJvW acknowledges support from the ERC Starting Grant ClusterWeb 804208.
We thank Nissim Kanekar for help with GMRT exposure time calculations,
and Huib Intema for help with the SPAM pipeline. We thank the staff of
the GMRT that made these observations possible. GMRT is run by the
National Centre for Radio Astrophysics of the Tata Institute of
Fundamental Research. Some of the data presented in this paper were
obtained from the Mikulski Archive for Space Telescopes (MAST) at the
Space Telescope Science Institute. The specific observations analyzed
can be accessed via
\dataset[10.17909/d922-3v11]{https://doi.org/10.17909/d922-3v11}.

\facilities{Keck:II (Deimos), HST (ACS), GMRT, XMM} 

\software{SAS (v19.0.0; Gabriel et al. 2004), mc3gmm code (Golovich et
  al. 2019), FIATMAP code (Wittman et al. 2006), SExtractor (Bertin \&
  Arnouts 1996)}

\vskip1.5cm 
\bibliography{ms}

\end{document}